\documentclass[a4paper,11pt]{article}
\usepackage{pos}

\title{Tracker alignment in CMS$:$
interplay with pixel local reconstruction}

\author*[a]{Ana Ventura Barroso}
\author{for the CMS Collaboration}

\affiliation[a]{Deutsches Elektronen-Synchrotron\\
  Notkestraße 85, 22607 Hamburg, Germany}

\emailAdd{ana.ventura.barroso@desy.de}

\abstract{The CMS silicon tracking system measures the trajectories of charged particles with a hit resolution of the order of microns in the pixel detector and tens of microns in the strip detector. One of the most important inputs for track reconstruction is the precision with which the tracker geometry is known. Therefore the position, orientation, and curvature of each tracker sensor must be precisely determined. Changes in the operating conditions can cause movements in the different substructures and also in the sensors. For maintaining the targeted precision, frequent corrections are needed, and the procedure to determine these corrections is commonly referred to as tracker alignment. Due to accumulated radiation during data taking, the response of the sensors changes over time. This affects the local reconstruction of pixel hits and consequently the result of the alignment procedure. In this contribution, the alignment procedure in CMS is introduced, as well as the dedicated calibration for the pixel local reconstruction. The effect of the change in the local reconstruction due to aging of the sensors on the alignment procedure is discussed.}

\FullConference{%
  10th International Workshop on Semiconductor Pixel Detectors for Particles and Imaging (Pixel2022)\\
  12-16 December 2022\\
  Santa Fe, New Mexico, USA}

\begin{document}
\maketitle

\section{The CMS tracker}

The CMS tracker is the largest fully silicon-based detector in the world in terms of number of sensors and total area. It comprises 1856 silicon pixel detector modules and 15148 silicon strip detector modules \cite{one}. In this publication the focus will be on the pixel detector, which is the closest subdetector to the interaction point. Due to its proximity to the beam, it is the most sensitive to radiation damage and requires the most precise alignment. It is composed of a barrel region (BPIX) and forward endcaps (FPIX).\\

The CMS pixel tracker was upgraded during the first Long Shutdown (LS1) in 2016-2017, referred to as the Phase-1 upgrade \cite{CMS-TDR-011}. It consists of four concentric barrel layers (L1-L4) at radii of 29, 68, 109, and 160 mm with in total 1184 modules, and three disks (D1-D3) on each end at distances of 291, 396, and 516 mm from the center of the detector, comprising 672 modules.\\

The pixel detector is located in a harsh radiation environment and the current data taking period, Run 3, is expected to double the collected integrated luminosity with respect to Run 2 (2015-2018). Therefore, during the second Long Shutdown (LS2), from 2019 to 2021, the Phase-1 pixel detector was refurbished \cite{JINST-16-P02027} to extend its lifetime while ensuring a performance in optimal conditions during the new data-taking period.
The pixel detector was extracted from the CMS experimental cavern and kept cold. The innermost layer in BPIX (L1) was completely replaced. Damaged modules were replaced (mostly modules of layer 2) and power supplies were upgraded from 600 to 800\,V, among other updates.  The pixel detector was reinstalled in 2021.\\

The tracker needs to provide excellent tracking performance to comply with the ambitious physics program of CMS. To reach this performance, it is crucial to know the absolute coordinates of the silicon sensors in the global CMS coordinate system with high precision. During the installation, the precision of the mechanical alignment is of the order of $0.1$ mm, which is larger than the design local hit reconstruction of the modules, $\mathcal{O}$(10$\mu$m). To achieve this precision, track-based alignment procedures are used.

\section{Tracker Alignment}
Tracker alignment is the procedure in which new parameters of the tracker geometry are determined. A big challenge is to obtain alignment corrections to a precision that ensures a good track reconstruction performance. Therefore, the goal is to determine with $\mathcal{O}$($\mu$m) the position, orientation, and surface deformation of all the modules in the pixel detector. This constitutes a major challenge due to the amount of degrees of freedom \cite{CMS-TRK-20-001}.\\

For every hit measurement $i$, position coordinates and errors are estimated within the local coordinate frame of the module. For every track $j$, hits assembled to tracks by the pattern recognition procedure get assigned a track parameter $q$ (e.g. parameters related to the track curvature and deflection by multiple scattering) \cite{TrckP}.
This depends strongly on the alignment parameters $p$, also called alignables or module parameters. The hit residual is defined as the difference between the measured hit position and the estimated intersection point of the particle trajectory and the two-dimensional module's plane.
The $\chi^2$-value reflecting the goodness of the track fit is given by the sum of all residual contributions normalised to their uncertainties of all hits associated to the track. In the presence of misalignment the value of the residual increases, thus increasing the $\chi^2$ value.\\

During operation the tracker needs to be realigned frequently due to changes in running conditions, such as fluctuations in temperature or in the magnetic field. For instance, magnet cycles (ramping up and down the magnet for maintenance reasons) can cause movements of the high level structures (half-barrels and half-disks) up to $\mathcal{O}$(mm).
The tracker is cooled during data taking, however cooling may be interrupted for maintenance purposes. Movements of the modules of $\mathcal{O}$(10 $\mu$m) after temperature variations have been observed.\\

In addition, the modules' performance is affected over time by the radiation dose received during operation, known as ageing of the modules. This effect produces a change of the Lorentz drift, which plays an important role in the pixel local reconstruction and the alignment calibration.

\section{Interplay with pixel local reconstruction}

Charge carriers traversing the silicon orthogonally to the direction of the magnetic field are deflected by an angle with respect to the electric field. This angle, which is known as Lorentz angle $\theta_{LA}$, depends on the electric field, the mobility of the charge carriers, and  the thickness of the active area. The measured Lorentz angle is used in the data reconstruction process to obtain an optimal position resolution and to minimize the potential bias in the hit reconstruction.
Hit pixels are combined to form clusters from neighbouring pixels. The charge measured within the clusters corresponds to the charge deposited by a single charged particle.\\

The Lorentz drift is not constant over time, due to radiation damage. The Lorentz angle effect causes degradation of the hit position resolution due to the increased cluster size. Furthermore, the fact that the cluster width is extended only in one direction causes systematic shifts of the hit position in the direction of the Lorentz angle.\\

The charge efficiency loss caused by radiation damage can be corrected by periodic recalibration and increase of the sensor bias voltage \cite{CMS-DP-2022-067}.
However, beyond a certain irradiation level full charge collection can not be recovered, leading to a degraded position resolution. Some residual effects caused by changes of the Lorentz angle are absorbed by the alignment procedure.\\

The sign of the Lorentz angle shift depends on the orientation of the electric field, so that the shift in the hit position in modules pointing inward is opposite with respect to this shift in outward pointing modules.
BPIX modules are arranged in ladders, where inward and outward modules can be aligned independently and thus their alignment can absorb the Lorentz angle effect.

\section{Monitoring tracking performance}
\subsection{Prompt Calibration Loop}
During data taking, the pixel detector operating configurations change over time, therefore a new set of alignment constants is needed periodically. This is achieved by defining intervals of validity (IOV) for a set of alignment constants.\\

To account for shifts in the different components of the pixel detector during data taking an automated alignment workflow is used. It provides an update of the alignment parameters within 48~hours. This workflow runs the MillePede-II \cite{MP} algorithm at Tier-0 as part of the Prompt Calibration Loop (PCL) and produces an alignment only of the pixel detector (without performing an alignment on the strip detector).\\

The alignment routine used during Run 2 and at the beginning of Run 3 performs a track-based alignment at the level of half barrels and half cylinders (high level structures), with a total of 36 alignment parameters. It is known as Low Granularity Prompt Calibration Loop (LG PCL).\\

In Run 3 the accumulated radiation will increase considerably with respect to Run 2, affecting the Lorentz drift in the tracker modules. This effect can be absorbed by re-aligning as often as possible the modules and by increasing the granularity of the alignment.
Therefore, for Run 3 the High Granularity Prompt Calibration Loop (HG PCL) has been deployed after a commissioning period. It is a track-based alignment as the LG PCL, but at the level of smaller support structures (ladders and panels), which increases the number of alignment parameters to $\sim$ 5k.
The usage of the HG PCL also replaces the need of manual HG alignments after new pixel calibrations.\\

\subsection{Distribution of Median of Residuals}
In the following figures, we compare the performance of three tracker geometries for Run 3:
\begin{itemize}
\item Alignment during data taking (black): alignment constants provided by the automated alignment, which runs online as part of the PCL. In the LG PCL alignment configuration used for the first period of data taking, corresponding to $\sim $11 $\mathrm{fb}^{-1}$, before the technical stop, the pixel detector is aligned at the level of half barrels and half cylinders.

\item Mid-year re-reconstruction (red): alignment corrections for the first portion of the 2022 data, corresponding to $\sim $9 $\mathrm{fb}^{-1}$, derived at the level of single modules for the pixel detector and strip subdetector, using 120M collision tracks recorded during pp collision runs at $\sqrt{s}$ = 13.6\,TeV and 8.5M cosmic ray tracks collected at 3.8 T magnetic field. Alignment constants for the last $\sim $2 $\mathrm{fb}^{-1}$ before the technical stop were provided by the automated alignment run offline after data taking in the High Granularity Prompt Calibration Loop configuration.

\item End-of-year re-reconstruction (blue): alignment constants for the data taking period after the technical stop, corresponding to $\sim $30 $\mathrm{fb}^{-1}$, provided by the automated alignment in the HG PCL configuration running online as part of the PCL workflow.  The starting geometry after the technical stop does not correspond to the alignment constants of the last IOV of the mid-year re-reconstruction geometry but was provided by the HG PCL alignment as well.
\end{itemize}

\begin{figure}[h]
\includegraphics[width=14cm]{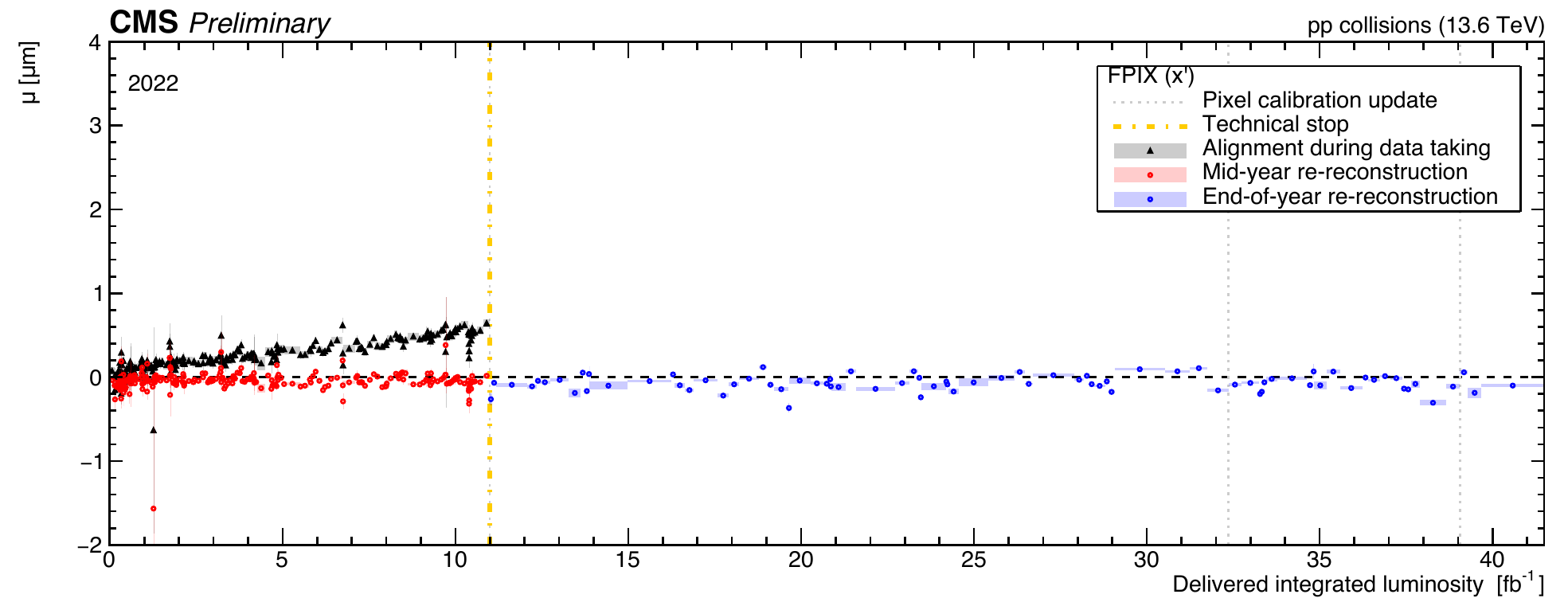}
\caption{The mean value of the distribution of median residuals is plotted for the local-$x$ ($x^\prime$) direction in the forward pixel detector (FPIX) as a function of the delivered integrated luminosity. The local $x$ axis is parallel to $\vec{E}$ $\times$ $\vec{B}$ where  $\vec{E}$ is the electric field of the sensor and $\vec{B}$ is defined to be the axial magnetic field \cite{CMS-TRK-20-001}. The vertical grey dotted lines indicate a change in the pixel tracker calibration and the yellow line a 4 week technical stop. The uncertainty corresponds to the standard mean error of the displayed quantity. Each color corresponds to a different alignment campaign \cite{CMS-DP-2022-070}.}
\centering
\label{FPIXmu}
\end{figure}

One of the tools used for monitoring the tracking performance is the Distributions of Medians of unbiased track-hit Residuals (DMR). Each track is refitted using the alignment constants under consideration, and the hit prediction for each module is obtained from all of the other track hits. The median of the distribution of unbiased track-hit residuals is then taken for each module and is added to a histogram. The width of the DMR constitutes a measure of the local precision of the alignment results, while for the mean value deviations from zero indicate possible biases. The width also has an intrinsic component due to the limited number of tracks, meaning that distributions can only be compared if they are produced requiring the same number of hits per module, as is the case for each set of figures shown.\\

The variable $ \mu$ is defined as the mean value (estimated by a Gaussian fit) of the distribution of the medians of the track-hit residuals computed per module in a given tracker substructure. The mean value corresponding to each IOV is extracted for the different alignment geometries and shown as a function of the delivered integrated luminosity.\\

The DMR for the FPIX is shown in Figure \ref{FPIXmu}. The online alignment with LG PCL at the beginning of data taking (black) deviates from zero due to changes in conditions during data taking. This deviation is corrected by the offline alignment after reprocessing (red). For the HG PCL the mean value of the distribution of median residuals is consistently closer to zero, showing improved stability with respect to the automated alignment in the LG PCL configuration.\\

\begin{figure}[h]
\includegraphics[width=7.5cm]{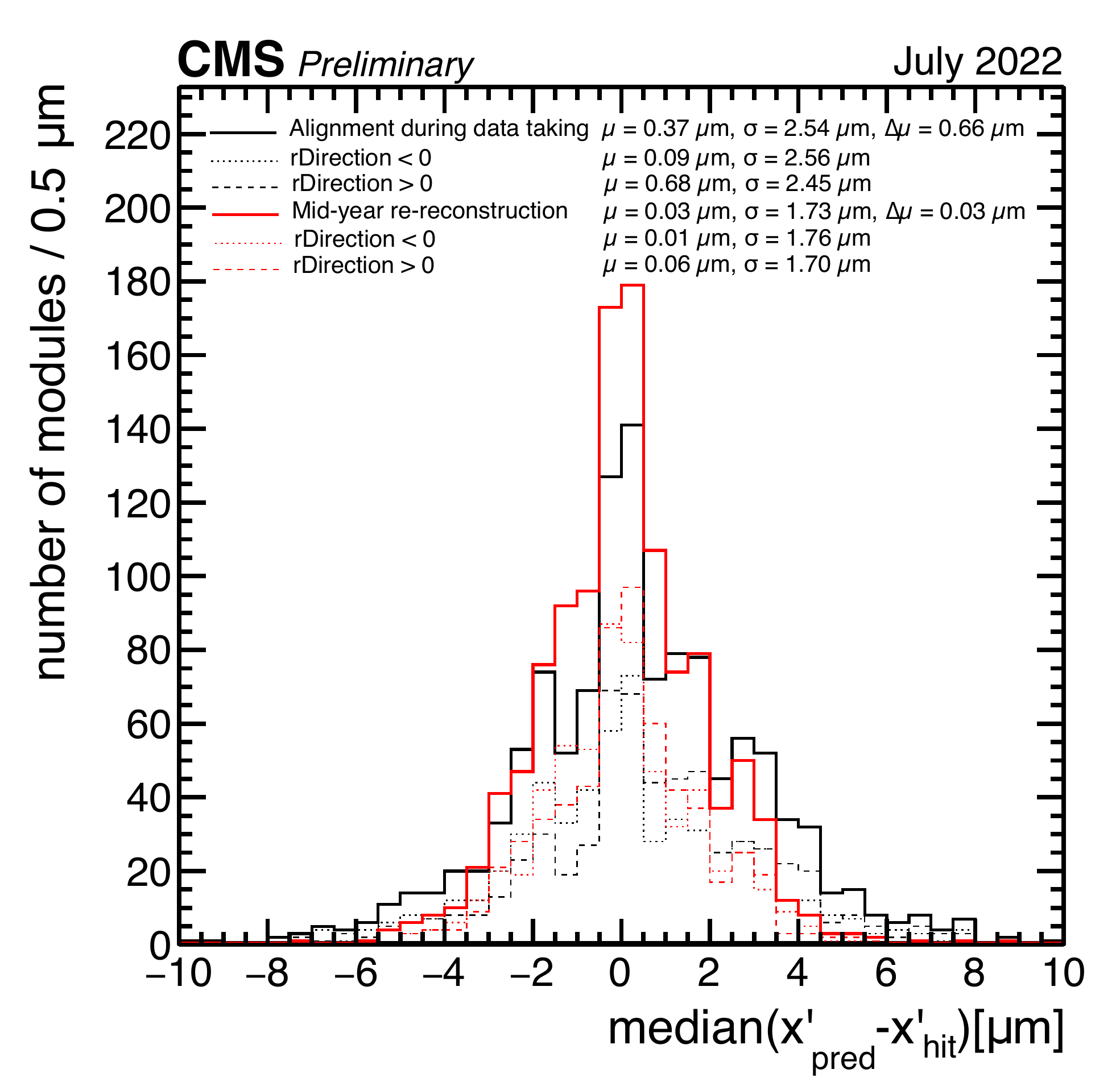}
\includegraphics[width=7.5cm]{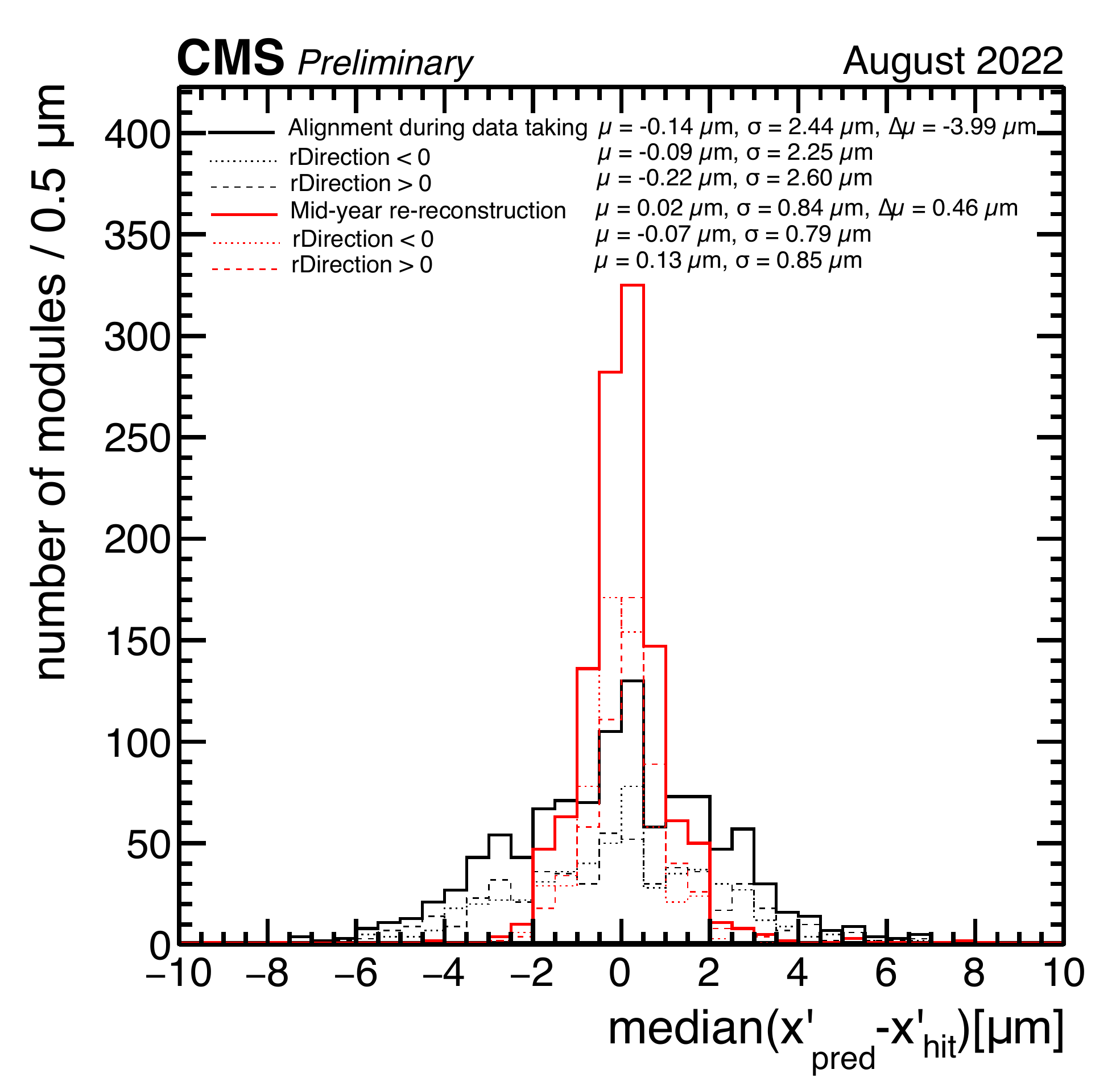}
 \caption{The distribution of median residuals is shown for the local $x^\prime$-direction in the barrel pixel detector for two different periods of time, namely during the months of July (left) and August (right) 2022 before the 4 week technical stop. The black line corresponds to the results provided by the automated alignment in the LG PCL. The red line indicates the geometry for the mid-year re-reconstruction. The quoted means $\mu$ and standard deviations $\sigma$ are the parameters of a Gaussian fit to the distributions, and $\Delta \mu$ denotes the difference of the mean values obtained separately for the modules with electric field pointing radially inwards and outwards in the local-$x$ ( $x^\prime$) direction \cite{CMS-DP-2022-070}.}
\centering
\label{DMR}
\end{figure}

As mentioned before, the sign of the Lorentz angle  shift depends on the orientation of the electric field. Therefore the shifts in the hit positions for inward and outward modules are opposite.
In the barrel region, DMR distributions can be obtained separately for the modules with electric field pointing radially inwards or outwards, as shown in Figure \ref{DMR}. A significant degradation of the alignment during data taking after 1 month of operation is observed. The mid-year re-reconstruction is capable of recovering from the change in conditions. The dashed lines show the DMR values for the inward and outward pointing modules for both alignment geometries. The difference of the mean values of the median residuals, $\Delta \mu = \mu_{\mathrm{inward}}-\mu_{\mathrm{outward}}$, constitutes an index of goodness in recovering Lorentz angle effects, where a mean value deviating from zero is a hint of residual biases due to the accumulated effects from radiation in the silicon sensors.

\begin{figure}[h]
\includegraphics[width=15cm]{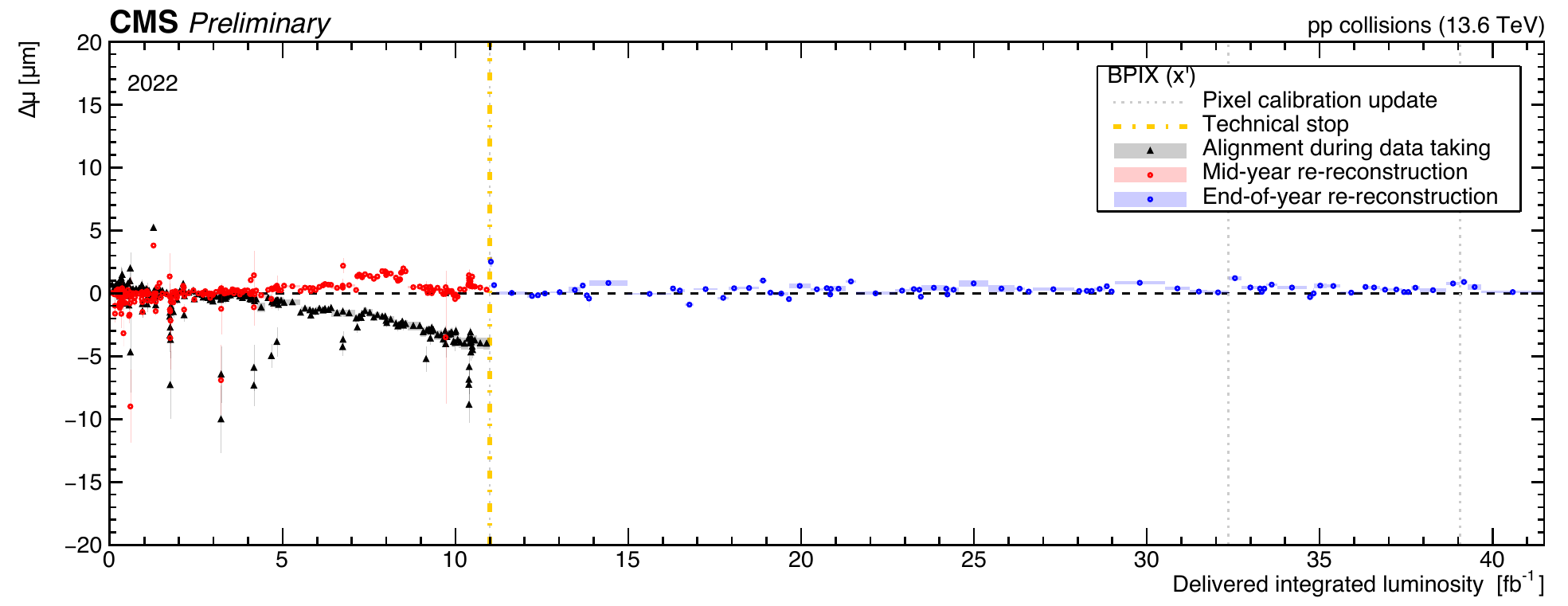}
\caption{$\Delta \mu$ is shown for the pixel barrel modules as a function of the delivered integrated luminosity. The uncertainty corresponds to the square root of the quadratic sum of the uncertainties calculated separately for the inward and outward pointing modules. The vertical grey dotted lines indicate a change in the pixel tracker calibration and the yellow line a 4 week technical stop. Each color corresponds to a different alignment campaign \cite{CMS-DP-2022-070}.}
\centering
\label{BPIXdm}
\end{figure}

Figure \ref{BPIXdm} shows the DMR of the $\Delta \mu$ for BPIX as a function of the delivered integrated luminosity. The online alignment with LG PCL at the beginning of data taking (black) and the offline alignment after reprocessing (red) deviate from zero due to shifts of the Lorentz angle caused by radiation damage. In the mid-year re-reconstruction an improvement of the difference of mean values is observed at around $\sim $ 9 $\mathrm{fb}^{-1}$, which corresponds to the start of the HG PCL offline alignment.
The online HG PCL (blue) corrects the position bias developed during data-taking and uncorrected by local reconstruction.\\

\begin{figure}[h]
\includegraphics[width=15cm]{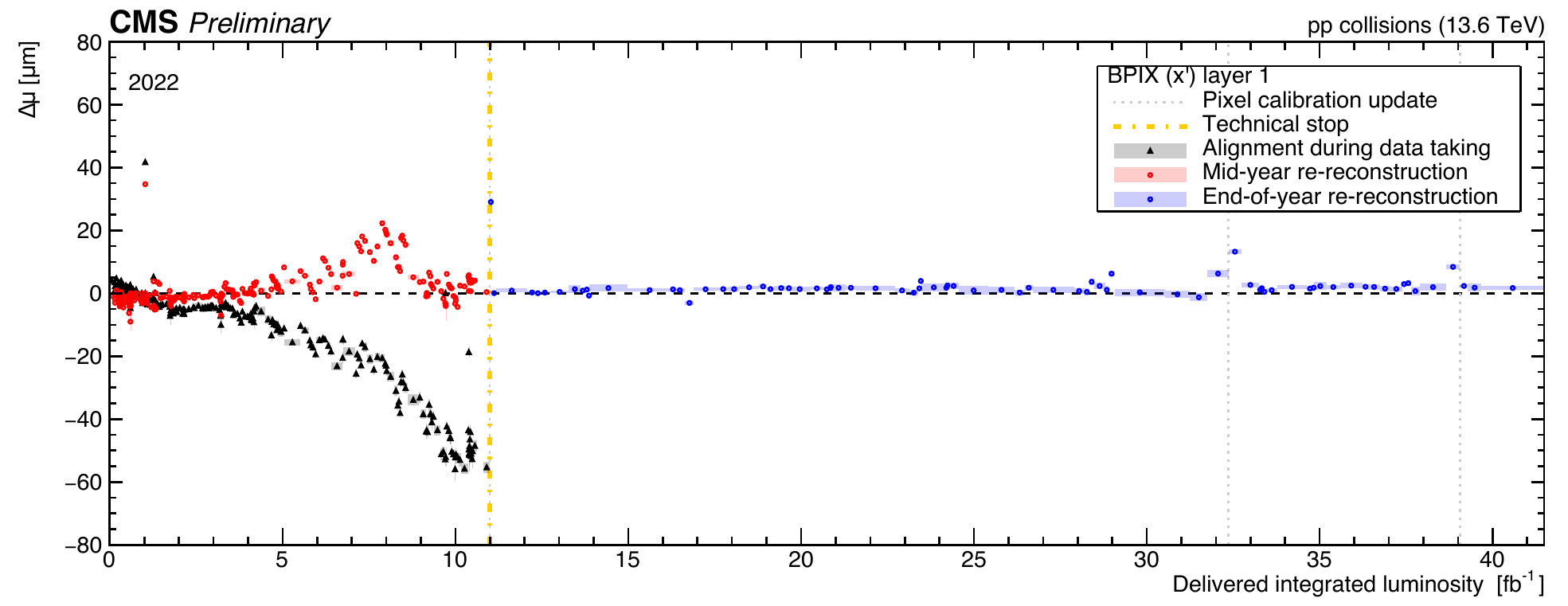}
\caption{$\Delta \mu$ is shown for the pixel barrel modules in layer 1 as a function of the delivered integrated luminosity. The uncertainty corresponds to the square root of the quadratic sum of the uncertainties calculated separately for the inward and outward pointing modules. The vertical grey dotted lines indicate a change in the pixel tracker calibration and the yellow line a 4 week technical stop. Each color corresponds to a different alignment campaign  \cite{CMS-DP-2022-070}.}
\centering
\label{DMUL1}
\end{figure}

In Figure \ref{DMUL1} the $\Delta \mu$ for BPIX layer 1 is shown. The radiation effects in layer 1 are more significant since it is the closest layer to the interaction point and therefore receives the most radiation damage. The mid-year re-reconstruction includes updates of the detector geometry with increased granularity than the alignment during data taking, which allows the effects of accumulating radiation damage to be mostly absorbed in the alignment procedure.\\

During the change in the pixel tracker calibration, the online HG PCL re-reconstruction needs some time to reabsorb the new conditions into the alignment (points around 32 $\mathrm{fb}^{-1}$ and 39 $\mathrm{fb}^{-1}$), but it quickly recovers.

\section{Summary}

The relevance of the interplay between pixel local reconstruction and tracker alignment has been presented. The monitoring of the aging and of the Lorentz angle effect in silicon modules  as a function of time using trends of distributions of the median of the residuals were reviewed. Finally, the HG PCL has proven to be extremely efficient at absorbing the effect of radiation damage, reducing the need for manual updates of the alignment conditions and improving the quality of the alignment in the prompt reconstruction. The online HG PCL shows stable performance in Run 3.

\end{document}